\documentclass{elsart}

\usepackage{amssymb}
\usepackage{amsmath}
\usepackage{graphicx}
\usepackage{epstopdf}
\usepackage{dcolumn}
\usepackage{bm}
\usepackage{color}
\usepackage{array}
\newcommand{\PreserveBackslash}[1]{\let\temp=\\#1\let\\=\temp}
\newcolumntype{C}[1]{>{\PreserveBackslash\centering}p{#1}}
\newcolumntype{R}[1]{>{\PreserveBackslash\raggedleft}p{#1}}
\newcolumntype{L}[1]{>{\PreserveBackslash\raggedright}p{#1}}

\begin{document}

\begin{frontmatter}

\title{Identify influential spreaders in complex networks, the role of neighborhood}

\author{Ying Liu$^1$$^,$$^2$$^,$$^3$},
\author{Ming Tang$^1$$^,$$^4$}\ead{tangminghuang521@hotmail.com},
\author{Tao Zhou$^1$$^,$$^4$},
\author{Younghae Do$^5$}

\address{$^1$ Web Sciences Center, University of Electronic Science and Technology of China, Chengdu 611731, China\\
$^2$ School of Computer Science, Southwest Petroleum University, Chengdu 610500, China\\
$^3$ Section for Science of Complex System, Medical University of Vienna, Vienna 1090, Austria\\
$^4$ Big Data Research Center, University of Electronic Science and Technology of China, Chengdu 611731, China\\
$^5$ Department of Mathematics, Kyungpook National University, Daegu 702-701, South Korea\\
}

\begin{abstract}
Identifying the most influential spreaders is an important issue in controlling the spreading processes in complex networks. Centrality measures are used to rank node influence in a spreading dynamics. Here we propose a node influence measure based on the centrality of a node and its neighbors' centrality, which we call the neighborhood centrality. By simulating the spreading processes in six real-world networks, we find that the neighborhood centrality greatly outperforms the basic centrality of a node such as the degree and coreness in ranking node influence and identifying the most influential spreaders. Interestingly, we discover a saturation effect in considering the neighborhood of a node, which is not the case of the larger the better. Specifically speaking, considering the 2-step neighborhood of nodes is a good choice that balances the cost and performance. If further step of neighborhood is taken into consideration, there is no obvious improvement and even decrease in the ranking performance. The saturation effect may be informative for studies that make use of the local structure of a node to determine its importance in the network.
\end{abstract}


\end{frontmatter}

\section{Introduction}
Identifying the most influential spreaders is an important step in promoting the adoption of new ideas, products and innovations, and preventing the epidemic disease from being pervasive. Centrality measures are used to rank the importance of node in a network, such as degree~\cite{freeman1978},closeness centrality~\cite{sabi1966}, betweenness centrality~\cite{freeman1977}, PageRank centrality~\cite{brin1998}, LeaderRank centrality~\cite{linyuan2006,li2014} eigenvector centrality~\cite{bonacich2001}, dynamic-sensitive centrality~\cite{klemm2012,lin2015} and coreness centrality~\cite{kitsak2010}. It contains the idea that there is a relationship between the topological position of a node in the network and its influence and capacity in a spreading dynamics. Among them, degree is the simplest way and is most widely used, but it is based only on the local connections of a node. Subsequent research pointed out that the coreness of node as identified by the $k$-shell decomposition~\cite{bolobas1984} is a more accurate way in ranking node influence~\cite{kitsak2010}, which has a low computational complexity of $O(N+E)$~\cite{batagelj2003}, where $N$ is the network size and $E$ is the number of edges. In addition, the $k_S$ index has a good robustness, which means that the relative ranking of the $k_S$ value for the same node remains unchanged when the network structure is incomplete, missing even up to 50\% of the edges~\cite{kitsak2010}.

It is pointed out that the importance of a node is not determined solely by its direct connections, but also depends on the connection of its neighbors~\cite{newman2010}. Research results show that ranking measures by considering the neighborhood of a node are more accurate in identifying the spreading influence of nodes~\cite{chen2012,Senpei2014,makse2015,huyanqing2015}. For example, by considering the number of neighbors within 4-step from the node, a local centrality measure is proposed which outperforms the node centrality of degree and betweenness~\cite{chen2012}. The sum of the coreness of the nearest neighbors of a node is a better indicator of node spreading influence than the coreness~\cite{bae2014}. In a ranking algorithm of iterative resource allocating, by considering the centrality of neighbors in a resource allocating process, the final resource a node obtained is used to ranking the spreading capability of the node, which shows a great improvement over degree, closeness, and betweenness~\cite{ren2014}. In addition, there are works based on counting the number of possible infection paths~\cite{bauer2012, lawyer2014} in the neighborhood to determine node influence.

Intuitively, the larger neighborhood is taken into consideration, the more accurate we can predict the spreading outcome of a node. However, in the research of spreading phenomena in social networks, such as the spread of smoking, alcohol consumption, happiness, health screening, it is discovered that on average there is a significant relationship in behaviors between a node and its direct neighbors (1-step neighbor), and up to the neighbors' neighbors' neighbors (3-step neighbor), which is called the three degree of separation~\cite{christ2012}. This implies a influence range from the spreading origin to the affected population. In addition,it is a challenging task to collect the complete network information in some real-world networks~\cite{Senpei2014}, due to the large amount and the temporal and spatial change of the data, such as Twitter and Facebook. Analyzing a large neighborhood seems unfeasible in such networks.

Given the effect of neighborhood, in this paper we first propose a neighborhood centrality based on the centrality of a node and its neighbors within multiple steps. Specifically, we study the performance of the neighborhood centrality when 1-4 steps of neighbors are considered. We find that in general the neighborhood centrality outperforms the centrality of the node. Furthermore, in most of the studied networks, considering the neighborhood within 2-step will result in a good neighborhood centrality. When the considered steps is greater than 2, the improvement of ranking accuracy is not obvious and even negative. This means a saturation effect of the neighborhood on a node. Discovering the saturation effect is meaningful in that when we consider the effect of neighborhood, taking the 2-step neighborhood into account will balance the cost and effect.
\section{Methods}
In this part, we first introduce the centrality measures of degree and coreness, which are used as the benchmark centrality. Then we propose the neighborhood centrality based on the degree or coreness of a node and its neighbors. Finally we describe the susceptible-infected-recovered (SIR) model used in the spreading process, and give a brief description of the data sets used in the study.
\subsection{Degree centrality}
 The degree centrality of a node $i$ is defined as $k_{i}=\sum_{j\in G\backslash i}a_{ij}$, where $j$ is a node in the network $G$, and $a_{ij}=1$ if there is a link between node $i$ and node $j$, otherwise $a_{ij}=0$. Degree is the simplest centrality measure in quantifying node importance. The larger the degree, the more 1-step neighbors the node is able to influence directly.
\subsection{Coreness centrality}
The coreness centrality of a node is obtained in the $k$-shell decomposition process. The $k$-shell decomposition method is used to decompose the network into hierarchically ordered shells from the core to the periphery. Initially, nodes with degree $k=1$ are removed from the network together with their links. After removing all nodes with $k=1$, there may appear some nodes with only one link left. We continue to remove these nodes until there is no node with one link left in the network. The removed nodes are assigned with an index $k_s=1$. Next, nodes with degree $k\leq2$ are removed in a similar way and assigned with an index $k_s=2$. The pruning process continues removing higher shells until all nodes are removed. As a result, each node is assigned with a $k_s$ value, which is called the coreness of a node. The coreness reflects the location importance of a node in the network. A large $k_s$ means a core position in the network, while a small $k_s$ defines the periphery a network. Coreness centrality is considered to be a better measure than degree to identify the influential spreaders in a network~\cite{kitsak2010,garas2010}.
\subsection{Neighborhood centrality}
We propose a new centrality measure that encodes the centrality of a node and its neighborhood. We consider that the importance of a node depends not only on its direct neighbors (1-step neighbors), but also on its 2-step and even more steps of neighbors. A 2-step neighbor of a node is a direct neighbor of 1-step neighbor. Similarly, a n-step neighbor is a direct neighbor of (n-1)-step neighbor. The more steps, the larger range of neighborhood is taken into consideration. Meanwhile, the weight of the neighbor's influence may be different. Intuitively, when an infection starts at a seed node, the probability of a neighbor being infected decreases with the increase of its distance from the spreading origin. That is the larger distance, the smaller effect the spreading origin may have on the neighbor. Based on these assumptions, we define the neighborhood centrality of node $i$ encoding the centrality of $i$ and its n-step neighbor as
\begin{equation}
C^n_i(\theta)=\theta_{i}+a\sum_{j\in\Gamma_{i}}\theta_{j}+a^2\sum_{l\in\Gamma_{j}\backslash i}\theta_{l}+a^3\sum_{m\in\Gamma_{l}\backslash j}\theta_{m}+...+a^n\sum_{s\in\Gamma_{s-1}\backslash x}\theta_{s},
\end{equation}
where $\theta$ is the benchmark centrality, $n$ is the step of neighbors taken into consideration, $a$ is an adjustable parameter that ranges in $[0,1]$, and $\Gamma_{i}$ is the set of nearest neighbor of node $i$. In the last item of the equation, $s$ is the direct neighbor of a node $o$ which is the $(s-1)$-step neighbor of the considered node $i$, while the slashed node $x$ is the direct neighbor of node $o$ but is the $(s-2)$-step neighbor of $i$. Here we use degree and coreness as the benchmark centrality, and consider a neighborhood of up to 4-step.
This equation means that the neighborhood centrality encodes the centrality of a node and its neighbors. In addition, the neighbors' effect decreases with the increase of its distance from the origin node. In this work, we first set $a=0.2$ in equation (1) and then discuss its impact on the performance of neighborhood centrality when $a$ varies.
\subsection{The SIR model}
 We use the susceptible-infected-recovered (SIR) spreading model~\cite{anderson1992,moreno2002} to simulate the spreading processes on networks and record the spreading efficiency for each node. We then compare the performance of the neighborhood centrality with degree and coreness in ranking the spreading efficiency of nodes and identifying influential spreaders. In the SIR model, a node has three possible states: $S$ (susceptible), $I$ (infected) and $R$ (recovered). Initially, a single node is infected and all others are susceptible. The infection spreads from the seed node to other nodes in the network through edges. At each time step, infected nodes contact all its neighbors, and then recover (change to $R$ state) with probability $\mu$.  For simplicity we set $\mu=1$. Recovered nodes will not be infected any more and remain the state of recovered until the spreading stops. Susceptible individuals become infected with probability $\lambda$ when they are contacted by an infected neighbor. The spreading process stops when there is no infected node in the network. The proportion of recovered nodes when spreading stops is considered as the spreading efficiency, or spreading capability, of the origin node. While the final infected population may vary when the infection probability $\lambda$ varies, authors of Ref.~\cite{liu2015} pointed out that the relative ranking of the nodes' spreading efficiency remains invariant under a wide range of infection probabilities. In our simulations, we chose an infection probability $\lambda>\lambda_c$, where the $\lambda_{c}=\langle k\rangle/(\langle k^{2}\rangle-\langle k\rangle)$ is the epidemic threshold of the network determined by the heterogenous mean-field method~\cite{castellano2010}. Under this infection probability, the final infected population is above 0 and reach a finite but small fraction of the network size, in the range of $1\%$-$20\%$~\cite{kitsak2010} for most nodes as spreading origins. We realize the spreading process for $100$ times and use the average spreading efficiency of a node as its spreading efficiency $M$. We also discuss the performance of our proposed method under a mediate range of the infection probability in our result part.
\subsection{Data sets}
We study the performance of the neighborhood centrality on six real-world networks. The six real-world networks studied are: (1) Email (e-mail network of University at Rovira i Virgili, URV) ~\cite{guimera2003}; (2) CA-Hep (Giant connected component of collaboration network of arXiv in high-energy physics theory)~\cite{leskovec2012}; (3) Hamster (friendships and family links between users of the website hamsterster.com)~\cite{hamster2014}; (4) PGP (an encrypted communication network)~\cite{boguna2004}; (5) Astro physics (collaboration network of astrophysics scientists)~\cite{newman2001}. (6) Router (the router level topology of the Internet, collected by the Rocketfuel Project)~\cite{spring2004}. The topological characteristics of the studied networks are listed in Table~\ref{basiccharacteristic}.
\begin{table}[htbp]
\begin{center}
\footnotesize
\caption{\label{basiccharacteristic}Topological characteristics of the real networks studied in this work. These characteristics include number of nodes ($N$), number of edges ($E$), average degree ($\langle k \rangle$), degree assortativity ($r$), clustering coefficient ($C$), epidemic threshold ($\lambda_{c}$), infection probability used in the SIR spreading in the main text ($\lambda$), average shortest distance $d$ of the network, and the average shortest distance from $R$ nodes to the spreading origin $i$ in a spreading process, then averaged over all nodes $i$, where $i$ is the top ranked 20\% nodes by spreading efficiency. We record this average distance as $d_R$.}

\begin{tabular}{ccccccccccccc}
\hline
Network & \textbf{$N$} & \textbf{$E$} & \textbf{$\langle k \rangle$} & \textbf{$r$} & \textbf{$C$} & \textbf{$\lambda_{c}$} & \textbf{$\lambda$} & \textbf{$d$} & \textbf{$d_R$} \\
\hline
Email  &1133 &5451 &9.6 &0.078 &0.220 &0.06 &0.08 &3.61 &2.22\\
CA-Hep &8638 &24806 &5.7 &0.239 &0.482 &0.08 &0.12 & 5.95 &2.93\\
Hamster  &2000 &16097 &16.1 &0.023 &0.540 &0.02 &0.04 &3.59 &2.04\\
PGP  &10680 &24340 &4.6 &0.240 &0.266 &0.06 &0.19 &7.49 &3.69\\
Astro  &14845 &119652 &16.1 &0.228 &0.670 &0.02 &0.05 &4.8 &2.95\\
Router  &5022 &6258 &2.5 &-0.138 &0.012 &0.08 &0.27 &6.45 &3.91\\
\hline
\end{tabular}
\end{center}
\end{table}

\section{Results}
We study the performance of neighborhood centrality in identifying influential spreaders by considering the node's neighbors of 1-step, 2-step, 3-step and 4-step, respectively, which are $C^1(\theta)$, $C^2(\theta)$, $C^3(\theta)$, $C^4(\theta)$ as defined in equation (1). We use the imprecision function proposed in ref.~\cite{kitsak2010} to quantify the performance of centrality measures in identifying influential spreaders. The imprecision function is defined as

\begin{equation}\label{imprecision}
\varepsilon(p)=1-\frac{M(p)}{M_{eff}(p)},
\end{equation}
where $p$ is the fraction of network size $N$ ($p\in[0,1]$). $M(p)$ is the average spreading efficiency of $pN$ nodes with the highest centrality, and $M_{eff}(p)$ is the average spreading efficiency of $pN$ nodes with the highest spreading efficiency. This function quantifies how close to the optimal spreading is the average spreading of the $pN$ nodes with the highest centrality. The smaller the $\varepsilon$ value, the more accurate the centrality is a measure to identify the most influential spreaders.

\begin{figure}
\begin{center}
\includegraphics[width=13.5cm]{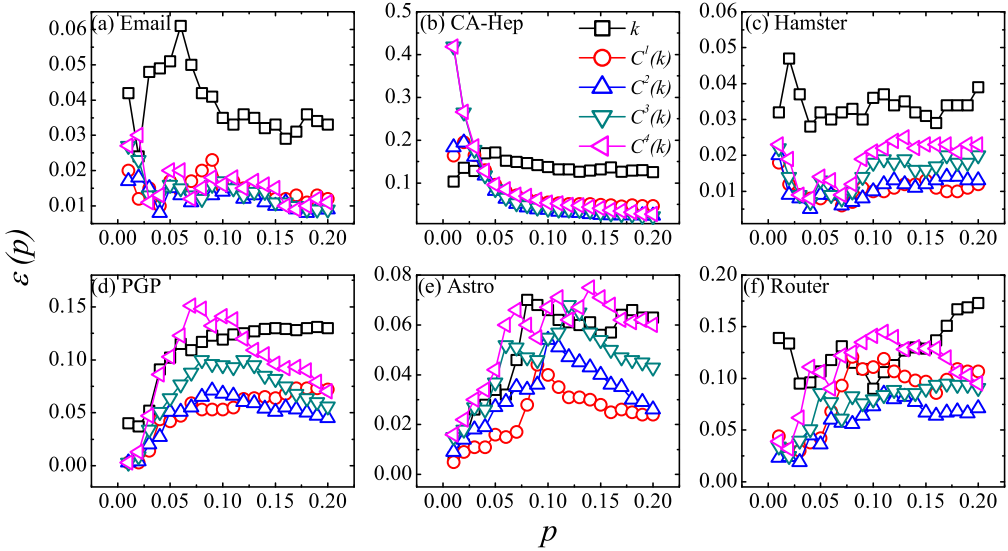}
\caption{The imprecision of centrality as a function of $p$ for six real-world networks. The imprecision of $k$ (black squares), $C^1(k)$ (red circles), $C^2(k)$ (blue uptriangles), $C^3(k)$ (green downtriangles), and $C^4(k)$ (purple lefttriangles) are compared in each network. $p$ is the proportion of nodes calculated, ranging from 0.01 to 0.2. In all studied networks, the lowest imprecision can be achieved within 2-step neighborhood.}
\label{figure1}
\end{center}
\end{figure}

\begin{figure}
\begin{center}
\includegraphics[width=13.5cm]{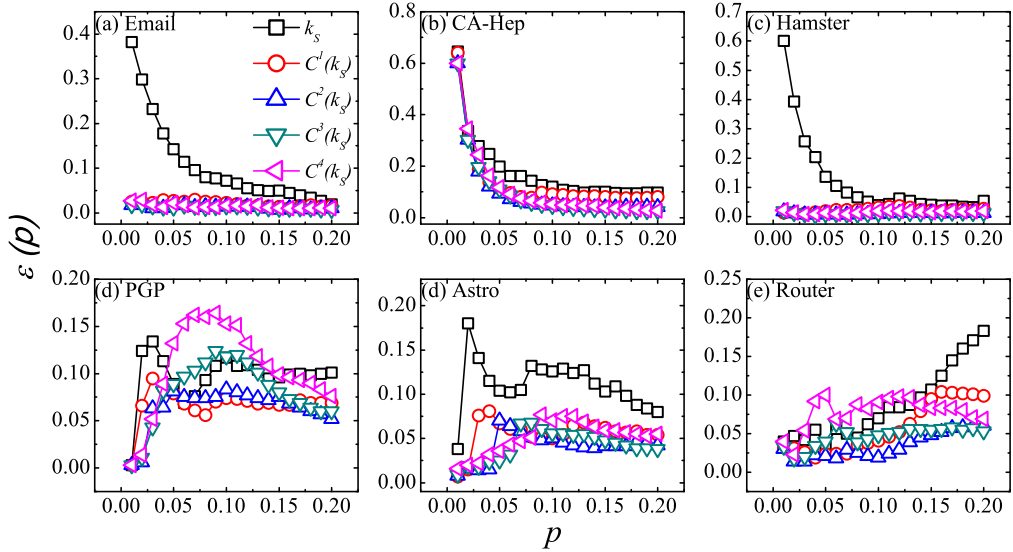}
\caption{The imprecision of centrality as a function of $p$ for six real-world networks. The imprecision of $k_S$ (black squares), $C^1(k_S)$ (red circles), $C^2(k_S)$(blue uptriangles), $C^3(k_S)$ (green downtriangles), and $C^4(k_S)$ (purple lefttriangles) are compared in each network. $p$ is the proportion of nodes calculated, ranging from 0.01 to 0.2. In all studied networks, the lowest imprecision can be achieved within 2-step neighborhood.}
\label{figure2}
\end{center}
\end{figure}

Here we compare the imprecision of $k$ with $C^1(k)$, $C^2(k)$, $C^3(k)$ and $C^4(k)$, as well as the imprecision of $k_S$ with $C^1(k_S)$, $C^2(k_S)$, $C^3(k_S)$ and $C^4(k_S)$, as shown in Fig.~\ref{figure1} and Fig.~\ref{figure2} respectively.
In Fig.1, we can see that for all studied networks, the lowest imprecision can be achieved at $C^1(k)$ and $C^2(k)$. In Email and CA-Hep, the neighborhood centrality outperforms the degree centrality at most $p$ values, and the imprecisions are very close under all steps considered. In Hamster and PGP, the imprecision of $C^1(k)$ and $C^2(k)$ are very close, and are lower than k, $C^3(k)$ and $C^4(k)$. In Astro the imprecision of $C^1(k)$ is the lowest, while in Router the imprecision of $C^2(k)$ is the lowest. In all, the $C(k)$ outperforms degree $k$, and a best neighborhood centrality is achieved when considering the neighbors of node in 1-step or 2-step for all the studied networks. When a larger step of 3 or 4 is considered, the performance of neighborhood even decreases. This demonstrates a saturation effect when considering the neighborhood of a node in determining its centrality.

When using the $k_S$ as the benchmark centrality, a similar saturation effect emerges as shown in Fig.~\ref{figure2}. In Email, CA and Hamster, the neighborhood centrality outperforms the coreness centrality, and are very close under all steps considered. In PGP, the imprecision of $C^1(k_S)$ and $C^2(k_S)$ are very close, and in general lower than that of $k_S$, $C^3(k_S)$ and $C^4(k_S)$. In Astro, the imprecision of $C^2(k_S)$, $C^3(k_S)$ and $C^4(k_S)$ are very close and smaller than $k_S$, except some p values. In Router, $C^2(k_S)$ and $C^3(k_S)$ have the lowest imprecision performance. In all, the $C(k_S)$ outperforms $k_S$ and a best performance can be achieved when considering the neighborhood within 2-step.

The imprecision function demonstrates the improved performance of neighborhood centrality in identifying the most influential spreaders. To make an explicit evaluation on the ranking capability of the topology-based neighborhood centrality on the spreading efficiency of nodes, we use the Kendall's tau correlation coefficient~\cite{kendall1938}. The Kendall's tau correlation coefficient is used to measure the ranking correlation of a same set in two ranking lists. It counts the number of concordant ranking pairs and the number of discordant ranking pairs in the two ranking lists. The correlation coefficient is defined as
\begin{equation}
\tau=\frac{\sum_{i<j}sgn[(x_{i}-x_{j})(y_{i}-y_{j})]}{\frac{1}{2}N(N-1)},
\end{equation}
where sgn(x) is a sign function, which returns 1 if $x>0$, -1 if $x<0$, and 0 if x=0. Here $N$ is the number of nodes in the list. $x_{i}$ and $x_{j}$ is the rank of node $i$ and node $j$ in ranking list 1, while $y_{i}$ and $y_{j}$ is the rank of node $i$ and node $j$ in ranking list 2. If node $i$ and node $j$ have a concordant rank in ranking list 1 and 2, $(x_{i}-x_{j})(y_{i}-y_{j})>0$. If node $i$ and node $j$ have a discordant rank in ranking list 1 and 2, $(x_{i}-x_{j})(y_{i}-y_{j})<0$. If node $i$ and node $j$ have a same rank in  either ranking list 1 or 2, $(x_{i}-x_{j})(y_{i}-y_{j})=0$. We take the topology-based ranking, that's the centrality measure, as ranking list 1 and the spreading-based ranking, that is the simulated spreading efficiency of nodes, as ranking list 2, and calculate the correlation coefficient. A large correlation coefficient implies a more concordant relation between the centrality and the spreading efficiency.

To make an explicit comparison, we calculate the improved tau ratio of neighborhood centrality over the benchmark centrality, which is
\begin{equation}
\eta_{\theta}=
 \begin{cases}
 \frac{\tau_{C(\theta)}-\tau_{\theta}}{\tau_{\theta}} &\tau_{\theta}>0\\
 \frac{\tau_{C(\theta)}-\tau_{\theta}}{-\tau_{\theta}} &\tau_{\theta}<0\\
 0 &\tau_{\theta}=0
 \end{cases},
\end{equation}
where $\theta$ is the benchmark centrality of $k$ and $k_S$, $\tau_{C(\theta)}$ is the correlation coefficient between the neighborhood centrality and spreading efficiency, $\tau_{\theta}$ is the correlation coefficient between the benchmark centrality and spreading efficiency, and $s$ is the number of steps of neighborhood that is taken into consideration. The improved tau ratio measures the increase of correlation for $C(k)$ over $k$ ($C(k_S)$ over $k_S$). As our main interest lies on the most influential spreaders, when we calculate the Kendall's tau correlation coefficient we only take the top ranked $pN$ of nodes by centrality into calculation.
\begin{figure}
\begin{center}
\includegraphics[width=13.5cm]{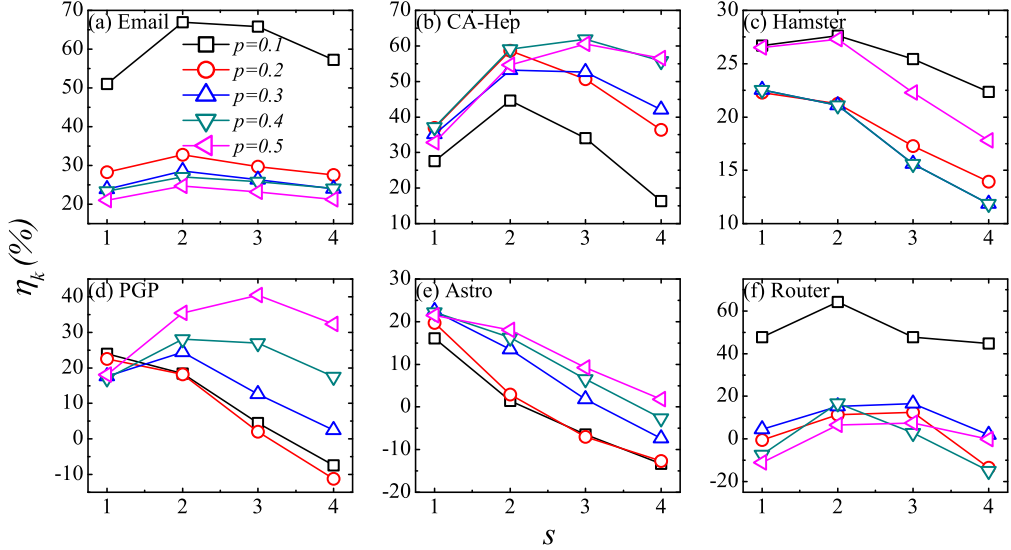}
\caption{The improved tau ratio $\eta_k$ of $C^s(k)$ over $k$ as a function of $s$ for six real-world networks. $s$ is the step of neighborhood considered. $p$ is the proportion of top ranked nodes by centrality that is calculated in Kendall tau's correlation coefficient, ranging from 0.1 to 0.5. In general, a largest increase in ranking correlation appears at 1-step or 2-step. Although in the networks of CA-Hep, PGP and Router there is some increase of $\eta_k$ at some p values for the 3-step, the increase is relatively small.}
\label{figure3}
\end{center}
\end{figure}
In Fig.~\ref{figure3}, the $\eta_k$ of $C(k)$ over $k$ for top ranked $pN$ nodes are demonstrated, where $p$ is in the range of $0.1$ to $0.5$ and the $C(k)$ is calculated from 1-step to 4-step of neighborhood. We can see that in general, the largest improved tau ratio is achieved within 2-step neighborhood. In Email, CA-Hep and Hamster, $\eta_k$ is greater than 0, and for most p values, the largest $\eta_k$ lies at 2-step. In PGP, the largest $\eta_k$ lies at 1-step when the top ranked 10\% and 20\% nodes are considered, while for other p values, the largest $\eta_k$ lies at 2-step or 3-step. In Astro, considering the 1-step neighbor will come to the largest $\eta_k$ for all p values. In Router, the largest $\eta_k$ lies at 2-step or 3-step for all p values. In all, by considering the neighborhood of 1-step or 2-step, the $C(k)$ has an increased ranking performance over $k$.

\begin{figure}
\begin{center}
\includegraphics[width=13.5cm]{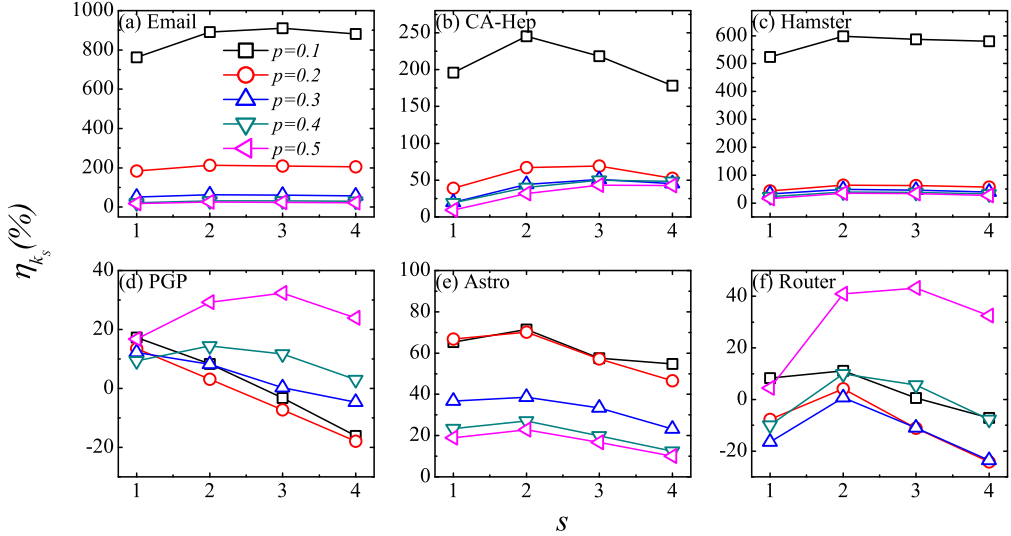}
\caption{The improved tau ratio $\eta_{k_S}$ of $C(k_S)$ over $k_S$ as a function of $s$ for six real-world networks. $s$ is the step of neighborhood considered. $p$ is the proportion of top ranked nodes by centrality that is calculated in Kendall tau's correlation coefficient, ranging from 0.1 to 0.5. In general, a largest increase in ranking correlation appears at 1-step or 2-step. Although in the networks of Email, PGP and Router there is some increase of $\eta_{k_S}$ at some p values for the 3-step, the increase is relatively small.}
\label{figure4}
\end{center}
\end{figure}

Similarly, the improved tau ratio $\eta_{k_S}$ of $C(k_S)$ over $k_S$ for top ranked $pN$ nodes are demonstrated in Fig. ~\ref{figure4}. In all networks except PGP, the largest $\eta_{k_S}$ lies at 2-step for most of the p values. In PGP the largest $\eta_{k_S}$ lies at 1-step for $p=0.1$ and $p=0.2$. When it comes to 3-step, the correlation decreases, which results in a negative $\eta_{k_S}$. It is worth noticing that in Email, CA-Hep and Hamster, the $\eta_{k_S}$ is very large in absolute value for $p=0.1$. As indicated in Ref.~\cite{liu2015,liu2015_2}, in networks of Email, CA-Hep and Hamster, the $k_S$ fails to identify the top ranked nodes in spreading efficiency. Thus the Kendall's tau correlation coefficient of $k_S$ and spreading efficiency for top ranked nodes in Email, CA-Hep and Hamster is very low, and is even negative for Email. In Email, the correlation coefficient of $k_S$ and node spreading efficiency for top 10\% nodes ranked by $k_S$ is $-0.08$, and the correlation coefficient of $C^1(k_S)$ and the node's spreading efficiency for top 10\% nodes ranked by $C^1(k_S)$ is $0.55$, thus $\eta_{k_S}\approx 788\%$ for $p=0.1$. In PGP and Router, the negative value of $\eta_{k_S}$ means that the ranking correlation of $C(k_S)$ is smaller than that of $k_S$.

In general, in all the studied six real-world networks, the neighborhood centrality outperforms the degree centrality and coreness centrality, and there exists a saturation effect when considering node's neighborhood, which mostly occurs at 2-step neighbors. As the above results are obtained when the using the parameter $a=0.2$ and the infection probability of $\lambda$, we test the performance under a range of the parameter $a$ and the infection probability respectively, and the results seems similar. 

\begin{figure}
\begin{center}
\includegraphics[width=13.5cm]{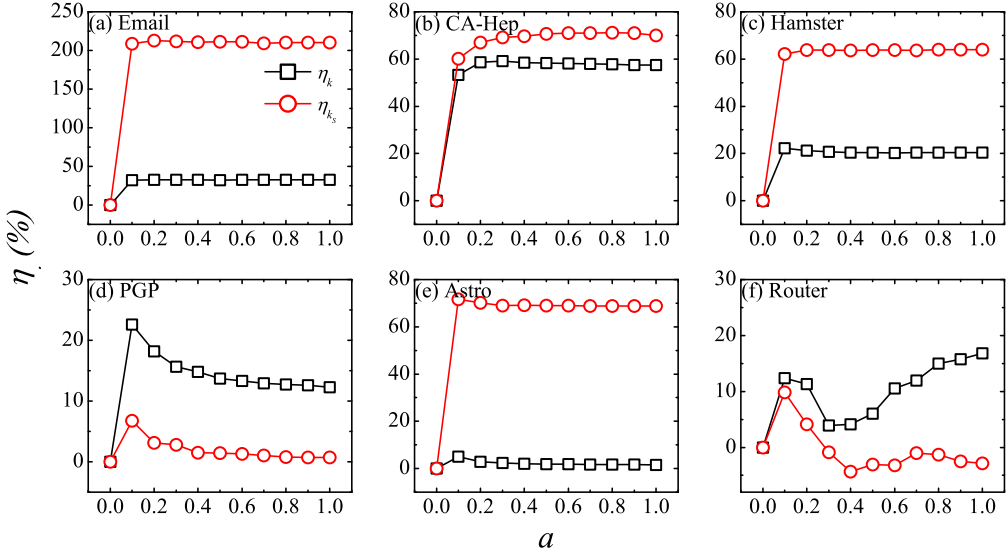}
\caption{The improved tau ratio $\eta_{k}$ ($\eta_{k_S}$) of $C^2(k)$ ($C^2(k_S)$) over $k$ ($k_S$) as a function of parameter $a$ in six real-world networks. The top ranked 20\% nodes by centrality is calculated in Kendall'tau correlation coefficient ($p=0.2$). In all networks except Router, $\eta_k$ and $\eta_{k_S}$ are relatively stable over all $a>0$. In Router, the $\eta_k$ is greater than 0 under all $a>0$. The $\eta_{k_S}$ is greater than 0 at $a=0.1$ and $a=0.2$ and then decrease to negative values. }
\label{figure5}
\end{center}
\end{figure}

\begin{figure}
\begin{center}
\includegraphics[width=13.5cm]{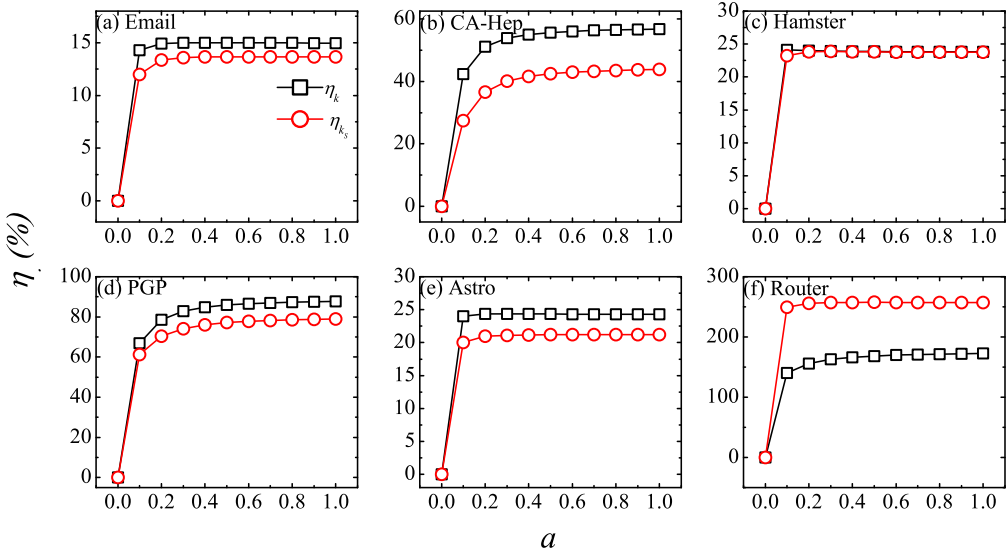}
\caption{The improved tau ratio $\eta_{k}$ ($\eta_{k_S}$) of $C^2(k)$ ($C^2(k_S)$) over $k$ ($k_S$) as a function of parameter $a$ in six real-world networks. All nodes in a network are calculated in Kendall'tau correlation coefficient ($p=1.0$). $\eta_k$ and $\eta_{k_S}$ are relatively stable under all $a>0$.}
\label{figure6}
\end{center}
\end{figure}

Now we concentrate on the $C^2(k)$ and $C^2(k_S)$, since in most cases they result in the best neighborhood centrality. We discuss the impact of the tunable parameter $a$ on the neighborhood centrality. We study the improved tau ratio of $C^2(k)$ over $k$ and  $C^2(k_S)$ over $k_S$ respectively at different $a$ values. As shown in Fig.~\ref{figure5}, we first focus on the top 20\% nodes ranked by neighborhood centrality. $a=0$ corresponds to $k$ or $k_S$. The $\eta$ in Email, CA-Hep, Hamster and Astro are stable under all $a>0$ values. In PGP, there is some fluctuation at a=0.2. As for Router, there is an obvious decrease at $a=0.4$, but the $\eta$ is always above 0. When all nodes of the network are taken into consideration, the improved tau ratio $\eta$ is very stable under all $a>0$ values, as shown in Fig.~\ref{figure6}. This implies that taking the neighborhood into consideration is quite influential, but the distance of neighbors is less important.

\begin{figure}
\begin{center}
\includegraphics[width=13.5cm]{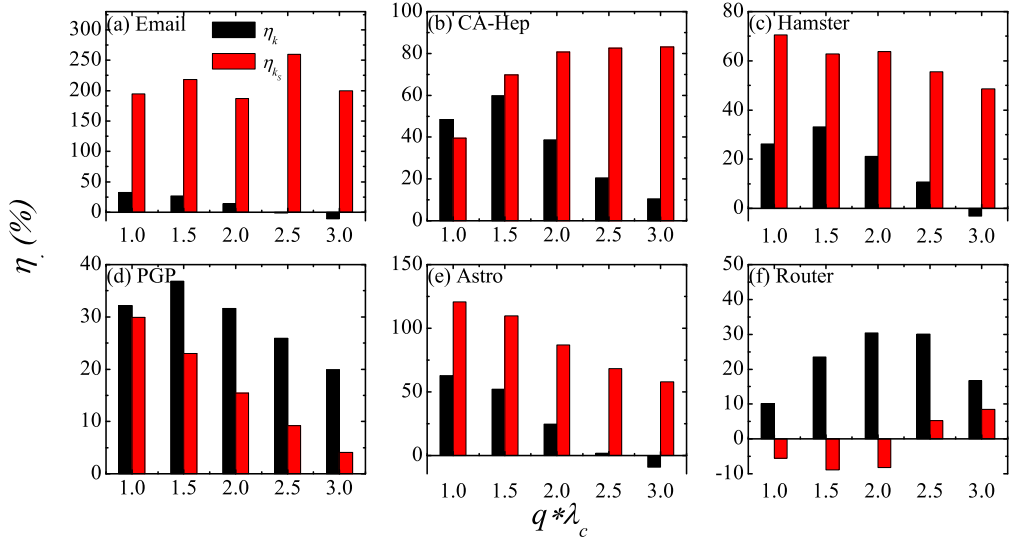}
\caption{The improved tau ratio $\eta_{k}$ ($\eta_{k_S}$) of $C^2(k)$ ($C^2(k_S)$) over $k$ ($k_S$) as a function of $q$ times of the epidemic threshold $\lambda_{c}$. $q$ ranges from 1.0 to 3.0. In most cased, the $\eta$ is greater than 0. There is a small negative value of $\eta_{k}$ at large $q$ values in Email, CA-Hep and Hamster, and a small negative value of $\eta_{k_S}$ at small $q$ values in Router.}
\label{figure7}
\end{center}
\end{figure}

\begin{figure}
\begin{center}
\includegraphics[width=13.5cm]{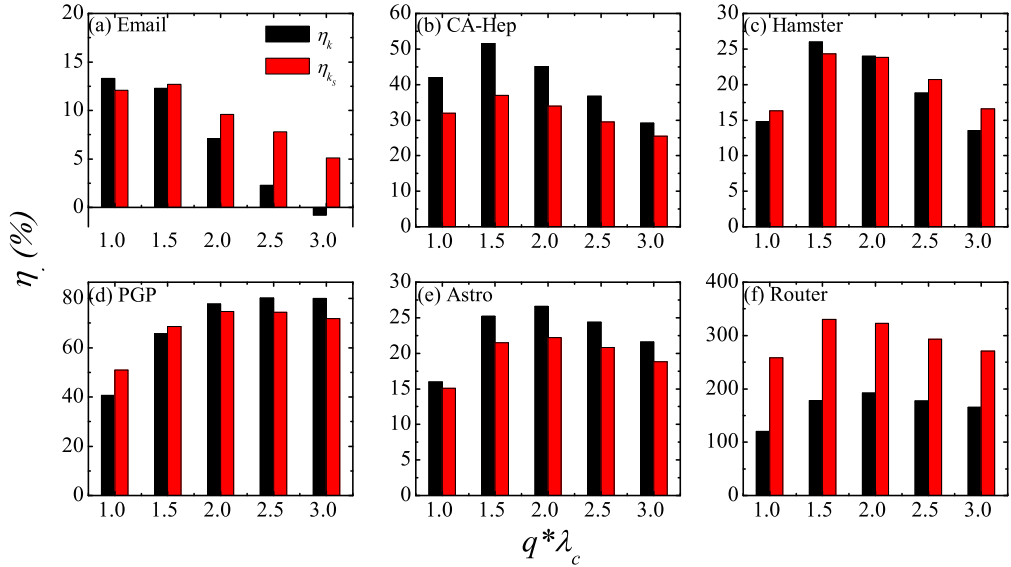}
\caption{The improved tau ratio $\eta_{k}$ ($\eta_{k_S}$) of $C^2(k)$ ($C^2(k_S)$) over $k$ ($k_S$) as a function of $q$ times of the epidemic threshold $\lambda_{c}$. $q$ ranges from 1.0 to 3.0. The improved tau ratio is greater than 0 in all studied networks and infection probability, except a small negative $\eta_k$ at three times the epidemic threshold in Email.}
\label{figure8}
\end{center}
\end{figure}

Finally, we move to explore the dependence of $\eta$ on the infection probability $\lambda$. We still focus on the $C^2(k)$ and $C^2(k_S)$. We present the $\eta_k$ and $\eta_{k_S}$ as a function of $q$ times of the epidemic threshold $\lambda_c$, where $q$ ranges from 1.0 to 3.0. As indicated by authors of Refs.~\cite{kitsak2010,liu2015}, the relative ranking of the spreading efficiency of nodes are not significantly influenced by the choice of infection probability in the spreading process. In addition, the infection probability should not be too large, otherwise the topological importance of the spreading origin is diminished, since under a large infection probability, nodes with low centrality value will be influential too, because there is a large chance that the disease is spread from the less influential spreaders to the more influential spreaders and then spread to the large part of the network~\cite{kitsak2010}. The result of considering the top 20\% nodes ranked by centrality is shown in Fig.~\ref{figure7}. In all networks except Router, the $\eta$ is greater than 0 at most infection rate. For Router, $\eta_k$ is greater than 0 under all infection probability. The $\eta_{k_S}$ is under 0 at some infection rate, but above 0 when $q=2.5$ and $q=3$. Although there is some fluctuation, the ranking correlation of $C(k)$ and $C(k_S)$ is higher than that of $k$ and $k_S$ in most cases. When all nodes of the network is taken into consideration, as shown in Fig.~\ref{figure8}, the $\eta$ is greater than 0 under all infection rates except a very small negative value at $3$ times of the $\lambda$ for $k$ in Email network. These results validate that the neighborhood centrality has a better performance than the benchmark centrality in a wide range of infection probability.

\section{Conclusion and Discussion}
Centrality measures are used in identifying influential spreaders. Here we propose a new centrality measure called neighborhood centrality, which encodes the centrality of a node and its neighbors. By simulating the SIR spreading process on six real-world networks, we find that the proposed neighborhood centrality outperforms the centrality of degree and coreness, which are two most widely used and simplest centrality measures, in identifying the influential spreaders. Furthermore, as we take the neighborhood of a node into consideration, we find that counter intuitively, it is not the case of the more the better. In most cases, considering the neighborhood of a node within 2-step will achieve a good performance while considering more steps of neighborhood will not obviously improve the performance or even result in some decrease. This demonstrates a saturation effect in considering the neighborhood, which coincides with the finding of "three degrees of separation" in social science~\cite{christ2012} in the sense that our way of considering the 2-step neighbors encodes the information of the 3-step neighbors. We also validate that the performance of the neighborhood centrality is stable under all values of the introduced parameter $a$ when $a>0$, as well as under a wide range of infection rates, which indicates a robustness of the proposed method.

The saturation effect of the neighborhood may be explained by the average shortest distance from a node to its potential infection range. We calculate the average shortest distance of the $R$ nodes to spreading origin, and average over the most influential nodes as the spreading origin, which are the top ranked 20\% nodes by spreading efficiency. This average shortest distance is recorded as $d_R$ and  listed in Table 1. We can see that $d_R$ is between 2 and 4. This implies that a node usually has an action scope. Considering neighbors within the scope will work, while considering neighbors out of this scope is less meaningful.

Many works make use of the neighborhood information to explore the structural and functional characteristics of nodes, such as decomposing and identifying the core-periphery structure of the network~\cite{garas2012,marius2013,rombach2014}, predicting missing links~\cite{zhou2009} and devising immunization strategies~\cite{salathe2010,yang2012,zhang2014}. We hope that the findings in this work will help to improve the researches by taking a suitable neighborhood range into consideration.

\section*{Acknowledgments}
This work was partially supported by the National Natural Science Foundation of China (Grant Nos. 11105025, 11575041, 61433014), the Scientific Research Starting Program of Southwest Petroleum University (No. 2014QHZ024), the Chinese Scholarship Council under No. 201406070071 and the National Research Foundation of Korea (NRF-2013R1A1A2010067).

\end{document}